# GPU-based finite-size pencil beam algorithm with 3D-density correction for radiotherapy dose calculation

**Xuejun Gu[1], Urszula Jelen[2], Jinsheng Li[3], Xun Jia[1], and Steve B. Jiang[1]**

[1]Center for Advanced Radiotherapy Technologies and Department of Radiation Oncology, University of California San Diego, La Jolla, CA 92037-0843, USA

[2]University Hospital Giessen and Marburg, Department of Radiotherapy and Radiation Oncology, Particle Therapy Center Marburg, 34045 Marburg, Germany

[3]Department of Radiation Oncology, Fox Chase Cancer Center, 333 Cottman Avenue, Philadelphia, PA 19111, USA

E-mail: sbjiang@ucsd.edu

Targeting at the development of an accurate and efficient dose calculation engine for online adaptive radiotherapy, we have implemented a finite size pencil beam (FSPB) algorithm with a 3D-density correction method on GPU. This new GPU-based dose engine is built on our previously published ultrafast FSPB computational framework [Gu et al. *Phys. Med. Biol.* **54** 6287-97, 2009]. Dosimetric evaluations against Monte Carlo dose calculations are conducted on 10 IMRT treatment plans (5 head-and-neck cases and 5 lung cases). For all cases, there is improvement with the 3D-density correction over the conventional FSPB algorithm and for most cases the improvement is significant. Regarding the efficiency, because of the appropriate arrangement of memory access and the usage of GPU intrinsic functions, the dose calculation for an IMRT plan can be accomplished well within 1 second (except for one case) with this new GPU-based FSPB algorithm. Compared to the previous GPU-based FSPB algorithm without 3D-density correction, this new algorithm, though slightly sacrificing the computational efficiency (~5-15% lower), has significantly improved the dose calculation accuracy, making it more suitable for online IMRT replanning.

**1. Introduction**

Online adaptive radiotherapy (ART) appears to be attractive, as it allows real-time adaptation of the treatment to daily anatomical variations (Wu *et al.*, 2002; Wu *et al.*, 2004; Court *et al.*, 2005; Mohan *et al.*, 2005; Court *et al.*, 2006; Wu *et al.*, 2008; Lu *et al.*, 2008; Ahunbay *et al.*, 2008; Fu *et al.*, 2009; Godley *et al.*, 2009; Men *et al.*, 2009; Gu *et al.*, 2009; Gu *et al.*, 2010b; Men *et al.*, 2010a; Men *et al.*, 2010b; Ahunbay *et al.*, 2010). However, it is challenging to implement online ART in clinical practice due to various technical barriers. One major barrier is to accurately compute dose distribution on the patient's new geometry in real time. Recently, a massive parallel computing architecture, graphics processing unit (GPU), has been introduced into the radiotherapy community and applied to accelerate computationally intensive tasks (Sharp *et al.*, 2007; Yan *et al.*, 2007; Li *et al.*, 2008; Samant *et al.*, 2008; Men *et al.*, 2009; Gu *et al.*, 2010b; Men *et al.*, 2010a; Men *et al.*, 2010b; Lu, 2010; Lu and Chen, 2010; Jia *et al.*, 2010b). Much effort has been devoted to utilize GPU to speed up dose calculation algorithms, including Monte Carlo (MC) simulation, superposition/convolution (S/C), and finite size pencil beam (FSPB) (Hissoiny *et al.*, 2009; Gu *et al.*, 2009; Jia *et al.*, 2010a; Jacques *et al.*, 2010; Hissoiny *et al.*, 2010).

The GPU-based FSPB model developed by our group is capable of calculating the dose distribution for a 9-field prostate treatment plan within 1 second (Gu *et al.*, 2009). However, like any other conventional FSPB models, our model only accounts for 1D density correction along the pencil beam depth direction and thus is less accurate when major inhomogeneities exist such as in lung cancer and head-and-neck cancer cases. Jelen and Alber (2007) have proposed a 3D density correction approach to improve the accuracy of an FSPB model (Jelen *et al.*, 2005). This improved FSPB model, termed as the *DC-FSPB* model in this paper, provides both lateral and longitudinal density corrections. Using a single flat $10 \times 10 \, \text{cm}^2$ beam in a lung case and a $6 \times 6 \, \text{cm}^2$ beam in a head-and-neck case, the authors initially demonstrated the accuracy of the model to be better than 2% for the majority of the voxels inside the field, which is a great improvement over the conventional FSPB models. In this paper, we will focus the implementation of this DC-FSPB model on GPU and exam its accuracy and efficiency using real clinical IMRT cases. We will 1) incorporate the DC-FSPB model into our GPU-based FSPB dose calculation framework; 2) systematically evaluate and demonstrate the accuracy improvement of the GPU-based DC-FSPB algorithm (*g-DC-FSPB*) over the GPU-based conventional FSPB algorithm (*g-FSPB*) under clinically realistic situations; 3) analyze in detail the ability of the g-DC-FSPB algorithm in handling various inhomogeneity situations; and 4) assess the efficiency of the g-DC-FSPB algorithm in comparison with the g-FSPB algorithm.

**2. Methods and Materials**

*2.1 An FSPB model with 3D density correction (DC-FSPB)*

77  In the DC-FSPB model proposed by Jelen and Alber (2007), the coefficients of the pencil
78  beam kernel were commissioned using the XVMC Monte Carlo simulation results
79  (Fippel *et al.*, 1999) in a homogenous water phantom and in a heterogeneous phantom
80  with slab geometry. Briefly, the dose at a spatial point **r** is the summation of the
81  contributions from all beamlets:

$$D(\mathbf{r}) = \sum D_i(\mathbf{r}) f_i ,$$  (1)

82  where $f_i$ denotes the photon fluence (or beamlet intensity) for the beamlet $i$. The dose
83  distribution of the beamlet $i$ with unit intensity from a point source located at $\mathbf{r_s}$ can be
84  formulated as:

$$D_i(\mathbf{r}) = F\big(x, y, \omega(\rho,t), u_x(\rho,t), u_y(\rho,t), x_0, y_0\big) \cdot A\big(t_{eq}, \theta\big) \cdot \big(\tfrac{SAD}{|\mathbf{r_a}|}\big)^2.$$  (2)

85  Here, $\mathbf{r_a}$ denotes the projection of the vector $\mathbf{r} - \mathbf{r_s}$ onto the beamlet direction. $x, y$ are
86  the projections of the vector $\mathbf{r} - \mathbf{r_s} - \mathbf{r_a}$ onto $x$-axis and $y$-axis of the plane perpendicular
87  to the beamlet direction. $x_0$ and $y_0$ represent the beamlet size. $x, y, x_0$, and $y_0$ are defined
88  at the isocenter plane. $SAD$ is the source to axis distance. $t$ is the portion of $|\mathbf{r_a}|$ below the
89  surface and $t_{eq}$ is the radiological depth. $\theta$ is the angle between the beamlet and its
90  corresponding beam central axis. $\omega$'s denote weighting factors and $u$'s are the steepness
91  parameters of the beam's penumbra. The function $F$ is the summation of two terms,
92  formulated as

$$F(x, y, \omega, u_x, u_y, x_0, y_0) = \sum_{i=1}^{2} \omega_i p(x, u_{ix}, x_0) p(y, u_{iy}, y_0).$$  (3)

93  Here, one term models the primary dose and the other one represents the secondary dose
94  accounting for scattering components. Each term is a product of two independent
95  exponential functions. Specifically $p(x, u_{ix}, x_0)$ is defined as:

$$p(x, u_{ix}, x_0) = \begin{cases} \sinh{(u_{ix} x_0)}\exp(u_{ix} x) & \text{for} \quad x < -x_0 \\ 1 - \cosh(u_{ix} x)\exp(-u_{ix} x_0) & \text{for} \quad -x_0 \le x \le x_0. \\ \sinh(u_{ix} x_0)\exp(-u_{ix} x) & \text{for} \quad x > x_0 \end{cases}$$  (4)

96  The term for $p(y, u_{iy}, y_0)$ is similarly defined. By adjusting the parameters in Eqs. (2)-
97  (4), we are able to shape the beamlet dose distribution in three dimensions. Along the
98  beamlet direction, $A(t_{eq}, \theta)$ is a function of radiological depth and off-axis angle, taking
99  care of heterogeneity correction along beamlet depth direction as well as the horn effect
100 at various off-axis distances. Perpendicular to the beamlet direction, the beam's
101 penumbra steepness is tuned according to local density $\rho$ as $u_1(\rho,t) = f_{u_1}(\rho) \cdot u_1^w(t)$
102 and a smoothed density $\hat{\rho}$ as $u_2(\rho,t) = f_{u_2}(\hat{\rho}) \cdot u_2^w(t)$, where the smoothed density $\hat{\rho}$ is
103 obtained by convolving the local density $\rho$ with a 3D symmetric Gaussian kernel. Here,
104 $u_1^w(t)$ and $u_2^w(t)$ are the parameters commissioned in a homogenous water phantom at a
105 geometrical depth $t$ and $f_{u_1}(\rho)$ and $f_{u_2}(\hat{\rho})$ are penumbra widening factors. The
106 weighting factors $\omega_i$ adjust the proportions of primary and secondary dose according to
107 the smoothed density $\hat{\rho}$ and the beamlet passing history using a formula $\omega_i(\rho,t) =$
108 $f_{\omega_i}(\hat{\rho}) \cdot \big(\omega_i^w(t) + \omega_i^{corr}(t)\big)$, where $f_{\omega_i}(\hat{\rho})$ adjusts weighting factors locally according
109 to a smoothed density $\hat{\rho}$. $\omega_i^w(t)$ is the commissioned weighting factor in a homogenous
110 water phantom at a depth $t$. $\omega_i^{corr}(t) = \int_0^t b\big(\rho(t')\big)dt'$, where $b\big(\rho(t')\big)$ is a parameter describing the changing of $\omega_i(\rho, t)$ values with the existence of heterogeneities. The details of the DC-FSPB model can be found in the reference (Jelen and Alber, 2007).

In this work, the model parameters were commissioned for the 6MV beam of a Varian 21EX linac using Monte Carlo simulated dose distributions. The dose distributions were calculated using the MCSIM Monte Carlo code (Ma *et al.*, 2002) together with a realistic source model (Jiang *et al.*, 2000) for a $10 \times 10$ cm$^2$ field with SAD=100 cm and SSD=90 cm. A slab geometry phantom of $30 \times 30 \times 30$ cm$^3$ dimension was used for commissioning. The slab of 15 cm thickness is inserted at 8cm below the phantom surface with the density varying from 0.1 to 2.0 g/cm$^3$. The parameters in the DC-FSPB model, such as $u, \omega, f(\cdot),$ and $b(\cdot)$, were obtained by fitting the dose distributions of the DC-FSPB model to those of the MCSIM simulation.

Once the parameters are established, the dose distribution for a board beam can be calculated using Eq. (1). Algorithm A1 given below illustrates the CPU implementation of the DC-FSPB algorithm. It, if skipping step 11, is degenerated to the FSPB algorithm with longitudinal density correction only.

**Algorithm A1:** An FSPB algorithm with 3D density correction implemented on CPU (DC-FSPB).

---

1.  Calculate a smoothed density distribution $\hat{\rho}$ by convolving the density distribution $\rho$ from patient CT data with a spherical Gaussian kernel;
2.  For each beamlet:
3.      Calculate the beamlet angle $\theta$;
4.      Extract the beamlet entrance and exit points on patient's body surface;
5.      Build a lookup table for radiological depth $t_{eq} = \int_0^t \frac{\mu(t)}{\mu_{H_2O}} dt'$;
6.      Build a lookup table for the weighting factor correction term:
$$\omega_i^{corr}(t) = \int_0^t b(\rho(t'))dt';$$
7.      For each voxel:
8.          For each beamlet such that the voxel is inside the region of interest (ROI)* of the beamlet
9.              Extract $A(t_{eq}, \theta)$ from the commissioned parameter lookup table;
10.             Extract $u_i^w(t)$ and $\omega_i^w(t)$ from the commissioned parameter lookup table;
11.             Calculate density corrected parameters:
$$u_1(\rho, t) = f_1(\rho)u_1^w(t);$$
$$u_2(\rho, t) = f_2(\hat{\rho})u_2^w(t);$$
$$\omega_i(\rho, t) = \omega_i^w(t) + \omega_i^{corr}(t) ;$$
12.             Calculate the dose according to Eqs. (1) and (2);
13.         End For
14.     End For
15. End For

---

152    *Here, ROI is defined as a cylinder of a radius of 5 cm centered at the beamlet
153    central axis.

154
155    *2.2 GPU implementation*

156
157    Algorithm A2 is the GPU implementation of Algorithm A1 using Compute Unified
158    Device Architecture (CUDA) programming environment. Similar to the CPU algorithm,
159    in Kernel 5, if we skip the density correction calculations, the g-DC-FSPB algorithm is
160    degenerated to the g-FSPB algorithm.

161    **Algorithm A2:** An FSPB algorithm with 3D density correction implemented on GPU
162    (g-DC-FSPB).

163    ────────────────────────────────────────────────────────────────

164    1.   Transfer the beam setup parameters, patient CT data, and commissioned model
165         parameters from CPU to GPU;
166    2.   Kernel 1: Perform an convolution to obtain smoothed density distribution $\hat{\rho}$ in
167         parallel (Step 1 in Algorithm A1);
168    3.   Kernel 2: Calculate the beamlet angle $\theta$ for all beamlets in parallel (Step 3 in
169         Algorithm A1);
170    4.   Kernel 3: Extract the beamlet entrance and exit points on the patient's body
171         surface for beamlets in parallel (Step 4 in Algorithm A1);
172    5.   Kernel 4: Build a radiological depth lookup table and a weighting factor
173         correction lookup table for all beamlets in parallel (Steps 5-6 in Algorithm A1);
174    6.   Kernel 5: Calculate dose to all voxels in parallel for all the beamlets (Steps 7-14
175         in Algorithm A1);
176    7.   Transfer the dose distribution from GPU to CPU.

177    ────────────────────────────────────────────────────────────────

178    The efficiency of a GPU code heavily relies on the efficiency of the memory
179    management. On a GPU card, available memory consists of constant memory, global
180    memory, shared memory, and texture memory. The constant memory is cached, which
181    requires only one memory instruction (4 clock cycles) to access. However, the available
182    constant memory is limited to 64 kB on a typical GPU card (such as NVIDIA Tesla
183    C1060). Due to the limited space, we store only those frequently accessed arrays with
184    constant values in the constant memory, such as the beam setup parameters and the
185    commissioned model parameters. The global memory is not cached and requires
186    coalesced memory access to achieve an optimal usage, but it has a large capacity (4GB
187    on one Tesla C1060 card) and is writable. Thus, we assign the radiological depth array
188    and the dose distribution array in the global memory since they requires memory writing.
189    The texture memory is read-only memory, but it is cached and the texture fetch are not
190    restricted by the coalescing memory access pattern to achieve high performance. The
191    density array is rested in the texture memory. By doing so, the performance is improved
192    with texture fetching in Kernel 1 and Kernel 4, where the convolution and integration
193    cannot follow the global memory coalescing accessing requirement.

194    The radiological depth and the weighting factor correction calculations require the
195 integration of the density functions along the beamlet direction, which is a
196 computationally intensive ray tracing problem. Siddon's algorithm is commonly used on
197 most CPU platforms for this task (Siddon, 1985). However, with Siddon's algorithm,
198 since the segment length that the beamlet central-axis intersects with each voxel is not
199 constant, the lookup table of the radiological depth (or the weighting factor correction
200 term) for each beamlet has to include two arrays: one storing the radiological depth (or
201 the weighting factor correction term) while the other auxiliary array listing the
202 corresponding geometrical depth. In Kernel 5 of Algorithm A2, for each voxel, we have
203 to search the geometrical depth array and then calculate the corresponding radiological
204 depth (or weighting factor). In order to reduce the memory usage and improve the
205 efficiency, in this work we adopt another approach to avoid the storage and search of the
206 geometrical depth auxiliary array. This approach computes the radiological depth and the
207 weighting factor correction term at the sampling points uniformly distributed along the
208 beamlet central-axis. The sampling step size is chosen as $d = \frac{1}{2}\min(\delta_x, \delta_y, \delta_z)$,
209 where $\delta_x$, $\delta_y$, $\delta_z$ represent the voxel size in $x, y$ and $z$ dimension. With this approach,
210 the storing and searching of the geometrical depth array becomes unnecessary. The
211 involved interpolation procedures can be conducted with high efficiency using the fast
212 on-chip linear interpolation function.

213    We compute the hyperbolic and exponential functions in Eq. (4) using CUDA
214 intrinsic function __expf($z$), which is about an order of magnitude faster than the standard
215 math function expf($z$). The maximum ulp (unit of least precision) error of __expf($z$) is
216 bounded by $2 + \text{floor}(\text{abs}(1.16 * z))$ (NVIDIA, 2010). For the data used in our g-DC-
217 FSPB model, since $z < 0.5$ the error of function __expf($z$) is actually bounded by 2
218 maximum ulp, which is equal to the error of the function expf($z$). Therefore, the use of
219 the intrinsic function __expf($z$) can greatly increase the efficiency without losing any
220 accuracy.
221
222 *2.3 Evaluation*
223

**Table 1**. Tumor site, number of beams, and case dimension for 5 head-and-neck (H1-H5) cases and 5 lung (L1-L5) cases.

| Case | Tumor Site | # of Beams | # of Beamlets | # of Voxels |
|---|---|---|---|---|
| H1 | Parotid | 8 (non-coplanar) | 7,264 | 128×128x72 |
| H2 | Hypopharynx | 7 (non-coplanar) | 4,429 | 128x128x72 |
| H3 | Nasal Cavity | 8 (non-coplanar) | 3,381 | 128x128x72 |
| H4 | Parotid | 5 (coplanar) | 4,179 | 128x128x72 |
| H5 | Larynx | 7 (non-coplanar) | 10,369 | 128x128x72 |
| L1 | Left lung, low lobe(close to pleura) | 6 (coplanar) | 637 | 128x128x80 |
| L2 | Right lung, low lobe (paravertebral) | 6 (coplanar) | 1,720 | 128x128x103 |
| L3 | Left lung, upper lobe (close to pleura) | 5 (coplanar) | 921 | 128x128x80 |
| L4 | Right lung, upper lobe (close to heart) | 7 (coplanar) | 841 | 128x128x80 |
| L5 | Left lung (middle) | 5 (coplanar) | 686 | 128x128x80 |

The g-DC-FSPB algorithm was evaluated for its accuracy against MCSIM algorithm (Ma *et al.*, 2002) and its efficiency using 10 real IMRT plans: 5 head-and-neck (H1-H5) cases and 5 lung (L1-L5) cases. All treatment plans were initially generated on the Eclipse planning system (Eclipse, Varian Medical Systems, Inc. Palo Alto, CA) and used to treat patients. Table 1 lists some relevant information for these 10 evaluation cases. The original CT images were down-sampled to the resolution of $0.4 \times 0.4 \times 0.25 \text{cm}^3$ for the dose calculations using MCSIM, g-FSPB, and g-DC-FSPB codes. Treatment plan parameters, including beam setup, leaf sequences, monitor units, *etc.*, were extracted from the Eclipse planning system and converted into RTP files as the input for MCSIM dose calculation. Leaf sequences and monitor units were reformatted into fluence map files as the input of g-FSPB and g-DC-FSPB codes. The resolution of the fluence maps (or the beamlet size) was selected as $0.2 \times 0.5 \text{ cm}^2$ with 0.2 cm along the MLC leaf motion direction.

For accuracy evaluation, the dose distributions calculated with MCSIM were used as the ground truth, with the maximum relative uncertainty less than 0.1% by simulating 2 billion particles for each beam. We computed the absolute dose in cGy for both g-DC-FSPB and MCSIM. The 3D $\gamma$-index distributions were computed using a GPU-based algorithm (Gu *et al.*). Dose distributions were evaluated with 3%-3mm criteria, where the 3% is relative to the maximum MCSIM dose value ($D_{max}$). The following statistical parameters were calculated and used as metrics to evaluate the dose calculation accuracy: 1) $\gamma^{max}$: the maximum $\gamma$ value of the entire dose distribution; 2) $\gamma_{50}^{avg}$: the average $\gamma$ values inside 50% isodose lines; 3) $P_{50}$: the percentage of voxels inside 50% isodose lines with $\gamma < 1.0$. For the efficiency evaluation, both g-FSPB and g-DC-FSPB dose calculations were conducted on an NVIDIA Tesla C1060 card. The data transferring time and the GPU computation time were recorded separately.

## 3. Results and Discussion

### 3.1 Accuracy evaluation

#### 3.1.1 Head-and-neck cases

**Table 2**. Gamma index evaluation results for 5 head-and-neck cases using the g-DC-FSPB algorithm. The corresponding g-FSPB results are given in parenthesis for comparison purpose.

| Case # | $\gamma^{max}$ | $\gamma_{50}^{avg}$ | $P_{50}$ |
|--------|----------------|---------------------|----------|
| H1 | 2.12 (2.16) | 0.30 (0.31) | 97.53% (97.32%) |
| H2 | 3.44 (4.11) | 0.28 (0.28) | 97.80% (97.01%) |
| H3 | 2.27 (2.36) | 0.46 (0.52) | 92.29% (86.39%) |
| H4 | 3.08 (3.11) | 0.61 (0.63) | 82.96% (81.56%) |
| H5 | 3.33 (3.37) | 0.61 (0.61) | 86.19% (86.09%) |

Table 2 summarizes the $\gamma$-index evaluation results for 5 head-and-neck cases. We can see that, for all 5 cases, $\gamma^{max}$ and $\gamma_{50}^{avg}$ values are smaller and $P_{50}$ values are larger for the g-

259   DC-FSPB algorithm, indicating that the new algorithm with 3D density correction
260   constantly outperforms the conventional FSPB algorithm. Specifically, we can put these
261   five cases into three scenarios:

262       *Scenario 1 (Case H1 and Case H2) - both g-FSPB and g-DC-FSPB algorithms are*
263   *accurate.* For these two cases, the average $\gamma$-index values are low (~0.3) and the passing
264   rates are high (>97%) for both the g-FSPB and g-DC-FSPB algorithms. By closely
265   inspecting the patient geometries and the treatment plans for Cases H1 and H2, we found
266   that there are only minor inhomogeneities on beams' paths and thus the g-FSPB
267   algorithm can calculate the dose distributions quite accurately. In such cases, there is not
268   much room for the g-DC-FSPB algorithm to improve the accuracy.

269       *Scenario 2 (Case H3) - the g-FSPB algorithm is less accurate but the g-DC-FSPB*
270   *algorithm can greatly improve the accuracy.* Figures 1(a), (b) and (c) show the dose
271   distributions for Case H3 calculated with the MCSIM, g-FSPB, and g-DC-FSPB
272   algorithms in the XY plane through isocenter, respectively. The $\gamma$-index distributions in
273   the same plane are presented in Figures 1(d) and (e), from which we can see that the $\gamma$-
274   index values decrease significantly at the nasal cavity region when the 3D density
275   correction is applied. The statistical analysis of the $\gamma$-index also shows that the g-DC-
276   FSPB dose distribution has a lower average $\gamma$-index value and a higher passing rate
277   compared to the g-FSPB result. These results indicate that the g-DC-FSPB algorithm is
278   capable of calculating dose more accurately in a low-density region (*e.g.* nasal cavity)
279   than the g-FSPB algorithm.

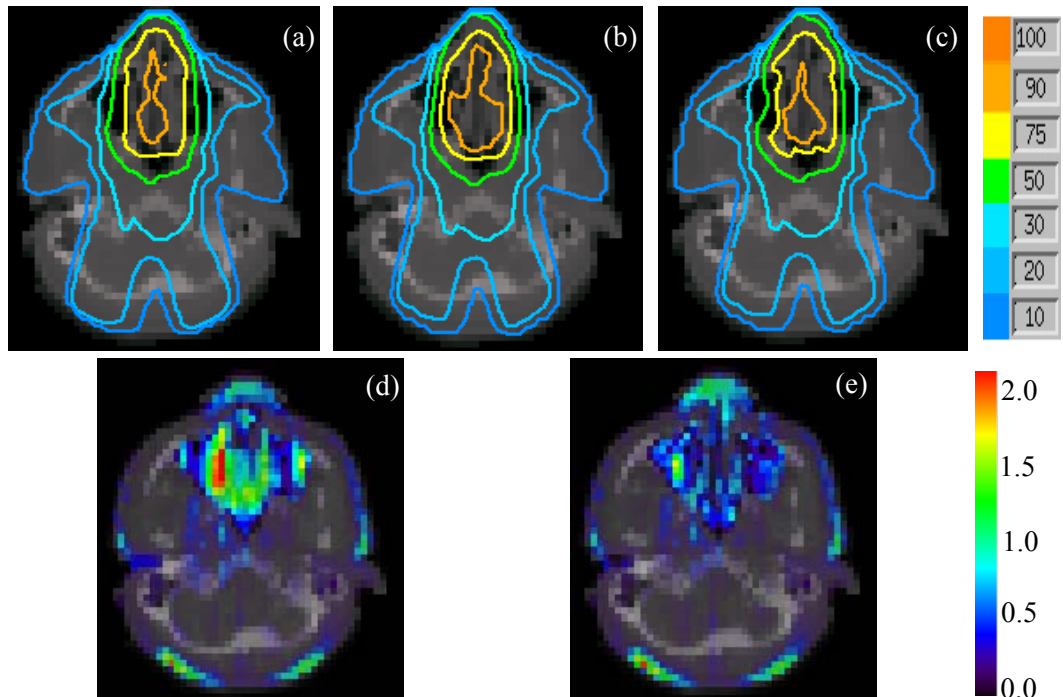

**Figure 1**. Dose distributions for Case H3 calculated with the MCSIM (a), g-FSPB (b), and g-DC-FSPB (c) algorithms in the XY plane through the isocenter. The $\gamma$-index distributions are shown in (d) for g-FSPB and (e) for g-DC-FSPB dose distributions in the same plane.

280

281        *Scenario 3 (Case H4 and Case H5) - both g-FSPB and g-DC-FSPB algorithms are*
282    *less accurate.* For these two cases, the g-FSPB dose distributions have large average $\gamma$-
283    index values ($\gamma_{50}^{avg}$~0.6) and low passing rates ($P_{50}$~86%). With 3D density correction,
284    the accuracy of the dose distributions is not much improved. By carefully inspecting
285    these two cases, we found that in both cases there are dental fillings of very high density
286    (~4.0 g/cm$^3$). Figure 2(a) shows dose distribution calculated with the g-DC-FSPB
287    algorithm and the density map of Case H4 in the XY plane through the isocenter, in
288    which we can clearly see the high density dental fillings. The dose difference maps
289    between the MCSIM and g-DC-FSPB dose distributions for each of the 5 co-planar
290    beams (309°, 0°, 51°, 102°, and 153°) are illustrated in Figures 2 (b)-(f). We can see that
291    the beam at angle 309° passes through the high density dental filling region before hitting
292    the target, causing a dose discrepancy up to 8% of $D_{max}$ between the g-DC-FSPB and
293    MCSIM results. This is because the density values near 4.0 g/cm$^3$ are far beyond our
294    commissioned density range and thus the g-DC-FSPB algorithm cannot find proper
295    parameters to accurately calculate the dose. For the other four beams, since they do not
296    pass through the high density region, the g-DC-FSPB dose distributions agree well
297    (within 1-2% of $D_{max}$) with the MCSIM dose distributions.

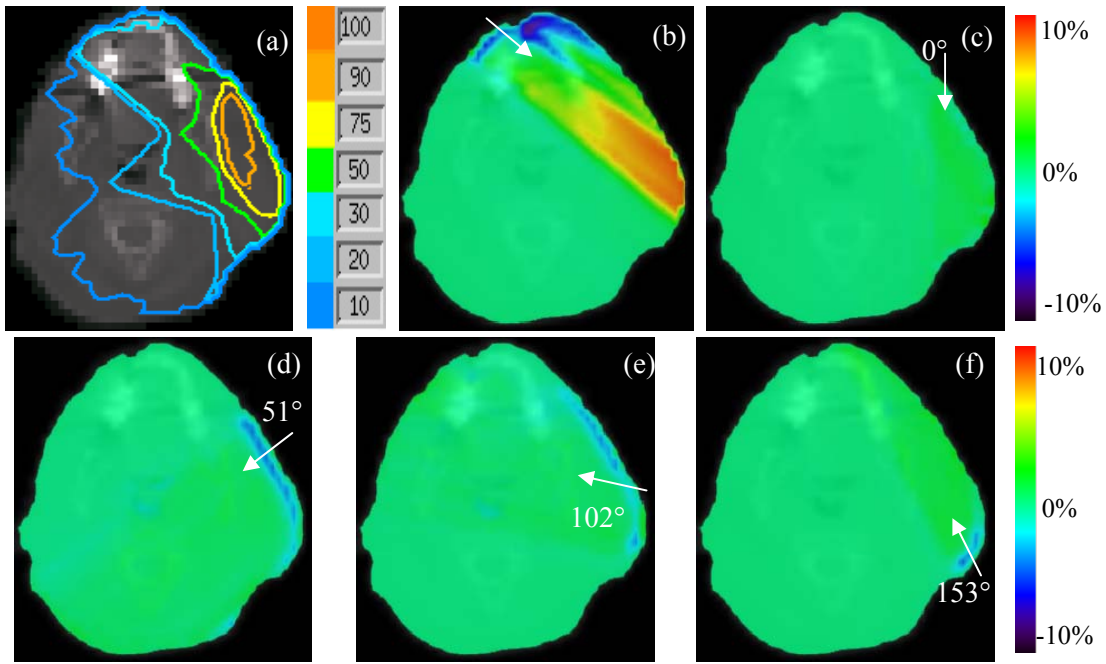

**Figure 2**. (a) Dose distributions for Case H4 calculated with the g-DC-FSPB algorithm in the XY plane through the isocenter. The dose difference maps in the unit of % $D_{max}$ between the g-DC-FSPB and the MCSIM results are shown in the same plane for each individual beam at the angle: (b) 309°, (c) 0°, (d) 51°, (e) 102°, and (f) 153°.

298
299    *3.1.2 Lung cases*
300
301        In Case L1, the tumor site is located in the lower lobe of the left lung, closing to the
302    pleura. The 5 out of total 6 beams do not pass through low-density lung regions before
303    hitting the target, in which cases the g-FSPB algorithm has sufficient accuracy. The last

304    beam goes through the low-density lung regions to reach the target and thus the 3D
305    density correction is needed to achieve high accuracy. The combined effect of all 6
306    beams is that, using the g-DC-FSPB method, $\gamma_{50}^{avg}$ is reduced from 0.45 to 0.24 and $P_{50}$
307    is increased from 94.81% to 99.35%.
308

**Table 3.** Gamma index evaluation results for 5 lung cases using the g-DC-FSPB algorithm. The corresponding g-FSPB results are given in parenthesis for comparison purpose.

| Case # | $\gamma^{max}$ | $\gamma_{50}^{avg}$ | $P_{50}$ |
|---|---|---|---|
| L1 | 1.53 (1.92) | 0.24 (0.45) | 99.35% (94.81%) |
| L2 | 2.35 (3.30) | 0.36 (0.71) | 96.64% (76.38%) |
| L3 | 1.68 (3.07) | 0.32 (0.75) | 99.16% (76.60%) |
| L4 | 2.70 (4.59) | 0.63 (1.53) | 81.33% (28.55%) |
| L5 | 2.19 (4.34) | 0.49 (1.13) | 90.24% (57.03%) |

309        In Case L2, the tumor site is close to the vertebral body. Three out of six beams strike
310    the targets without passing through low-density lung regions. For these beams the g-
311    FSPB algorithm can generate accurate results. For the other three beams, which pass
312    through lung areas before hitting the target, the g-FSPB algorithm becomes inadequate.
313    The dose distributions calculated with the MCSIM, g-FSPB and g-DC-FSPB algorithms
314    are plotted in the XY plane through isocenter in Figures 3(a), 3(b), and 3(c). In Figure
315    3(b), the g-FSPB dose is much higher than the MCSIM dose in the region indicated by
316    the arrow. When the g-DC-FSPB algorithm is used, this hot spot disappears, as shown in
317    Figure 3(c). Figures 3(d) and 3(e) plot the $\gamma$-index distributions in the same plane
318    calculated with the g-FSPB and g-DC-FSPB algorithms, respectively. The statistical
319    values shown in Table 3 also indicate a significant improvement of $\gamma_{50}^{avg}$ and $P_{50}$ values
320    using the 3D density correction method, where, $\gamma_{50}^{avg}$ values decrease from 0.71 to 0.36
321    and $P_{50}$ values increase from 76.38% to 96.64%. The situation for Case L3 is very similar
322    to that of Case L2.

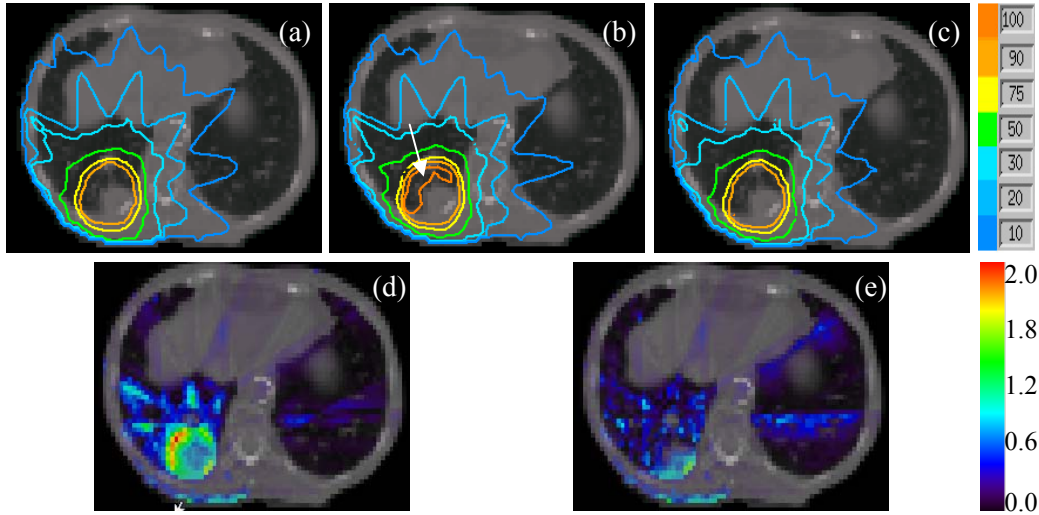

**Figure 3.** The dose distribution of Case L2 calculated with MCSIM (a), g-FSPB (b), and g-DC-FSPB (c) in the XY plane through the isocenter. The $\gamma$-index distributions in the same plane are illustrated in (d) for g-FSPB and (e) for g-DC-FSPB.

The tumor in Case L4 is in the middle of the lung, indicating that all beams have to pass through the low-density lung regions before hitting the target. The dose distributions in the XY plane through isocenter calculated with the MCSIM, g-FSPB and g-DC-FSPB algorithms are shown in Figures 4(a), (b), and (c). The $\gamma$-index distributions in the same plane calculated with the g-FSPB and g-DC-FSPB algorithms are plotted in Figures 4(d) and (e). From Figures 4(a), 4(b), and 4(d), we observe that in the high dose region, the g-FSPB algorithm heavily overestimates the calculated dose. From Figures 4(c) and 4(e), we can see that the g-DC-FSPB algorithm can correct the overestimation of the g-FSPB algorithm and greatly improve the agreement with MCSIM, especially inside the target region. However, in lung regions outside the target, the density correction is overdone, resulting in an underestimated dose.

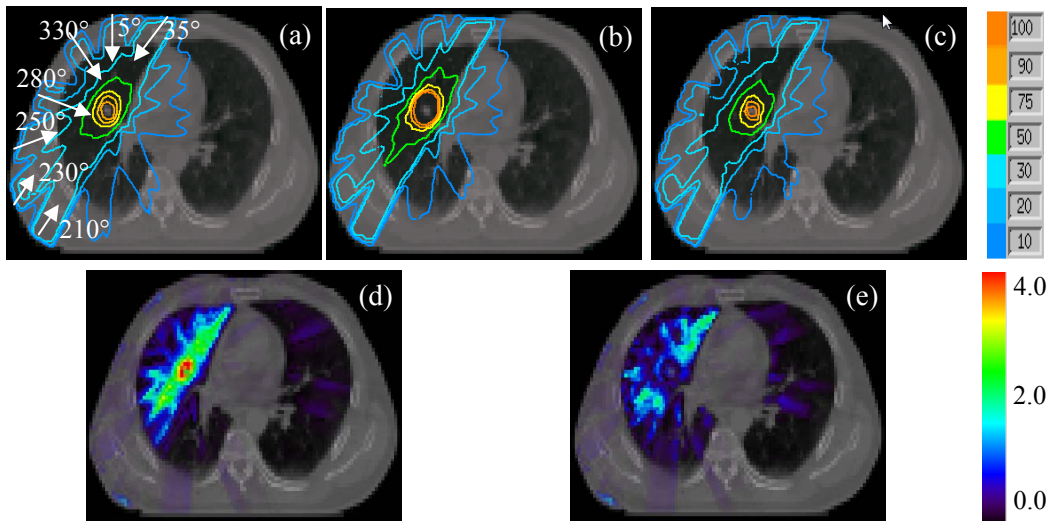

**Figure 4.** Dose distributions for Case L4 in the XY plane through the isocenter calculated with the MCSIM (a), g-FSPB (b), and g-DC-FSPB (c) algorithms. $\gamma$-index distributions for the g-FSPB (d) and g-DC-FSPB (e) algorithms are given in the same plane.

The dose distribution for Case L4 is analyzed individually for each of the 7 coplanar beams at the gantry angles of 35°, 5°, 330°, 280°, 250°, 230°, and 210°. In Figures 5, we plot the normalized depth dose curves and depth density curves for each beam. Here, we normalize three depth dose curves for each beam to the maximum dose calculated with MCSIM algorithm for that beam. We can see that, without 3D-density correction, all the depth dose curves exhibit a monotonic decrease after the maximum dose and do not show a clear inhomogeneity correction effect. In contrast, the depth dose curves calculated with the g-DC-FSPB algorithm exhibit a proper trend of density correction, *i.e.*, build-down and build-up effects, as indicated by the MCSIM depth dose curves. Overall, the calculated dose distribution of each beam is significantly improved with 3D-density correction. For the composite dose distribution, as shown in Table 3, the $\gamma$-index passing rate inside 50% isodose line has been improved from 28.55% to 81.33%. However, for some beams the g-DC-FSPB algorithm overcorrects the density effect, leading to a much underestimated dose in lung regions. This phenomenon is particularly obvious for gantry

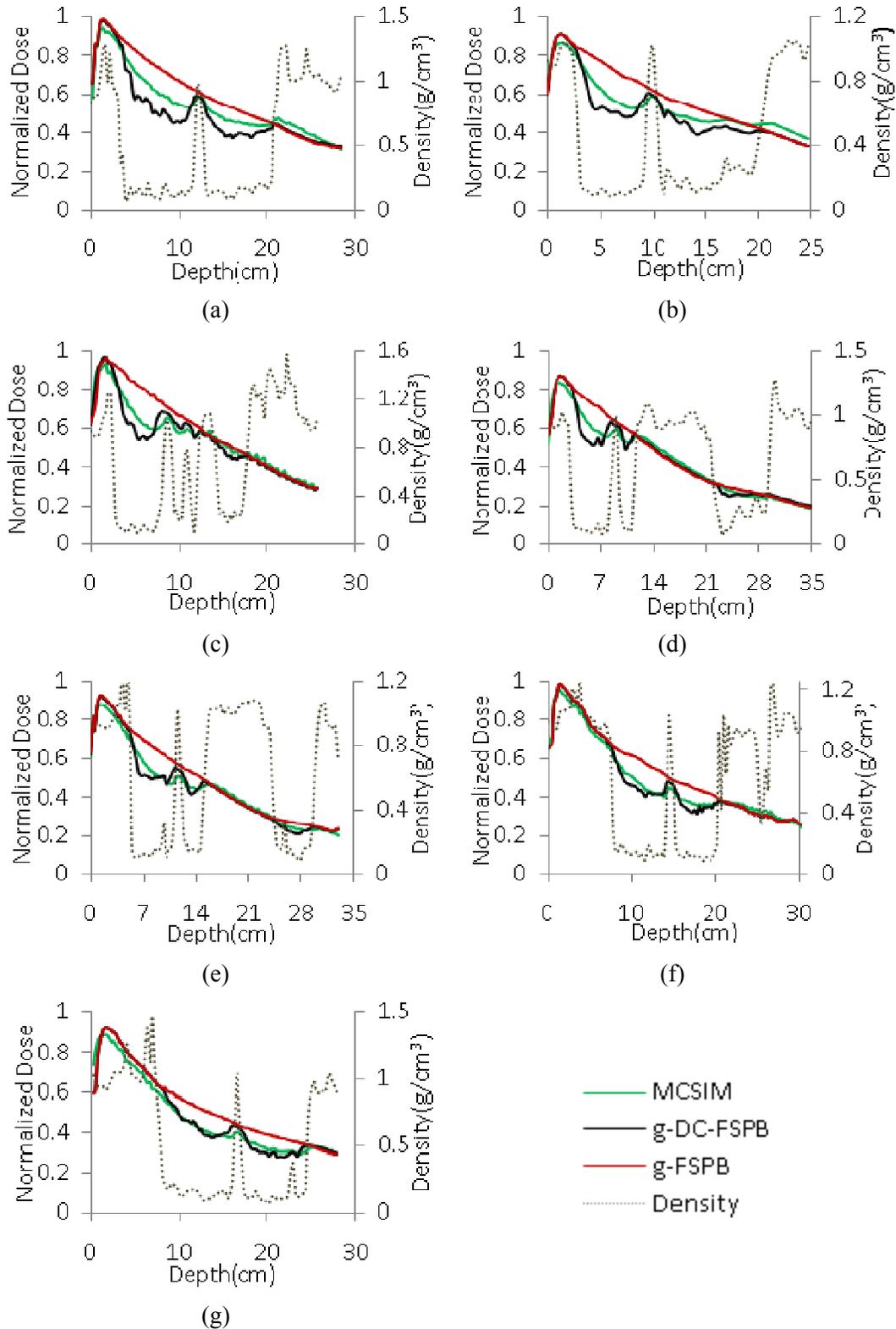

**Figure 5.** The depth dose curves and the depth density curves along the beam central axis for seven beams (a) 35°, (b) 5°, (c) 330°, (d) 280°, (e) 250°, (f) 230°, and (g) 210°. The depth dose curves are normalized to each beam's maximum dose calculated with MCSIM.

349  angle such 35°, which is mainly responsible for the discrepancy shown in Figure 4(e).
350  Similarly, for Case L5, the improvement of dose distribution achieved by the 3D-density
351  correction method is dramatic. However the γ-index passing rate in the 50% isodose line
352  for the g-DC-FSPB algorithm is still less satisfactory due to the similar overcorrection
353  issue.

354

355  *3.2 Efficiency evaluation*

356

357  Table 4 lists computation time for dose calculation using the g-FSPB and g-DC-FSPB
358  algorithms. We can see that the dose distribution of a realistic IMRT plan can be
359  computed at a very high efficiency. For 9 out of 10 testing cases, the dose calculation can
360  be completed within one second using either algorithm. For all 10 cases, the median data
361  transfer time between CPU and GPU is 0.2 seconds, and the median GPU computation
362  time is 0.37 seconds for the g-DC-FSPB algorithm and 0.33 seconds for the g-FSPB
363  algorithm. Since the computation time is so short, the data transfer time takes a
364  significant portion of the total computation time, up to 50% in Case L1. We can also see
365  that, while the accuracy of the g-DC-FSPB algorithm is much higher than that of the g-
366  FSPB algorithm, its efficiency sacrifice is quite mild (~5-15% slower in terms of the total
367  computation time).

368

**Table 4.** Dose calculation time using the g-FSPB (in parenthesis) and g-DC-FSPB algorithms for 10 testing cases. $T_{tr}$ is the data transfer time between CPU and GPU. $T_{gpu}$ is the GPU computation time. $T_{tot} = T_{tr} + T_{gpu}$.

| Case # | $T_{tr}$(sec) | $T_{gpu}$(sec) | $T_{tot}$(sec) |
|--------|---------------|----------------|----------------|
| H1     | 0.20          | 0.64 (0.55)    | 0.84 (0.75)    |
| H2     | 0.20          | 0.40 (0.35)    | 0.60 (0.55)    |
| H3     | 0.20          | 0.38 (0.34)    | 0.58 (0.54)    |
| H4     | 0.19          | 0.35 (0.32)    | 0.54 (0.51)    |
| H5     | 0.20          | 1.31 (1.10)    | 1.51 (1.30)    |
| L1     | 0.21          | 0.22 (0.20)    | 0.43 (0.41)    |
| L2     | 0.22          | 0.40 (0.36)    | 0.62 (0.58)    |
| L3     | 0.21          | 0.30 (0.25)    | 0.51 (0.46)    |
| L4     | 0.18          | 0.25 (0.23)    | 0.43 (0.41)    |
| L5     | 0.21          | 0.33 (0.29)    | 0.54 (0.50)    |
| Median | 0.20          | 0.37 (0.33)    | 0.56 (0.53)    |

369

370  **4. Conclusions**

371

372  In this paper, we detailed the implementation of the g-DC-FSPB algorithm. The
373  dosimetric evaluation of the g-DC-FSPB algorithm was conducted on 5 head-and-neck
374  and 5 lung IMRT treatment plans. Using the dose distributions computed with the
375  MCSIM Monte Carlo code as reference, we assessed the accuracy improvement of the g-
376  DC-FSPB algorithm over the g-FSPB algorithm.

For head and neck cases, 1) when only minor heterogeneities exist, the g-FSPB algorithm is already quite accurate and the improvement achieved by the g-DC-FSPB algorithm is mild; 2) when air cavities are near the target, the g-DC-FSPB algorithm can significantly improve the accuracy of dose distribution; 3) when there are high-density dental filling materials in the beam paths, the dose calculation accuracy of the g-DC-FSPB algorithm is unsatisfactory although there is still an improvement over the g-FSPB algorithm, due to the fact that such high-density materials were not considered in the commissioning process.

For all lung cases, the accuracy of calculated dose distributions is significantly improved with the 3D-density correction method. However, the degree of such improvement is highly dependent on inhomogeneities presented in the beam paths. When the majority of beams in a treatment plan reach the target without passing through the low-density lung region, the accuracy of dose distribution calculated by the g-FSPB algorithm is already satisfactory, while there is still a significant improvement with the 3D-density correction method. When more than half of the beams in a treatment plan have to pass through the low-density lung region before reaching the target, the accuracy of the g-FSPB algorithm is poor, while the g-DC-FSPB algorithm can dramatically improve the dose calculation accuracy.

In the original work of Jelen *et al.* (2007), better than 2% of accuracy was demonstrated for the majority of the voxels inside the field when using the DC-FSPB model, which seems better than our g-DC-FSPM algorithm. We would like to point out that their accuracy was accomplished for a single flat $10 \times 10 \text{cm}^2$ beam in a lung case and a $6 \times 6 \text{ cm}^2$ beam in a head-and-neck case, while our results were obtained for 10 real clinical IMRT cases.

Regarding the efficiency, we see that for 9 out of 10 testing cases, the dose calculation can be completed well within one second for both g-FSPB and g-DC-FSPB algorithms. The median GPU computation times are less than half a second for both algorithms. Compared to the g-FSPB algorithm, the g-DC-FSPB algorithm slightly sacrifices the computation efficiency, about 5-15% slower in terms of the total computation time. However, the significant accuracy improvement of the g-DC-FSPB algorithm far outweighs the slight efficiency lost, indicating that this algorithm is more suitable for online IMRT replanning.

**Acknowledgements**

This work is supported in part by the University of California Lab Fees Research Program and by an NIH/NCI grants 1F32 CA154045-01. We would like to thank NVIDIA for providing GPU cards for this project.